\documentclass[english]{article}
\usepackage[T1]{fontenc}
\usepackage[latin9]{inputenc}
\usepackage{babel}
\usepackage{mathrsfs}
\usepackage{amsmath}
\usepackage{amssymb}
\usepackage[unicode=true]
 {hyperref}
\begin{document}
\title{A note on a gap in the proof of the minimum distance for Projective
Reed-Muller Codes }
\author{ANDERS BJÆRT SØRENSEN}
\date{October 2023}
\maketitle

\section*{Introduction}

In {[}4, p.1569{]} I presented Theorem 1, which describes the basic
parameters dimension and minimum distance for projective Reed-Muller
codes (PRM codes). In {[}1, Remark 3.4{]} Prof. S. Ghorpade and R.
Ludhani identified a gap in my proof of this theorem, and subsequently
rectified it with ease, confirming the validity of the theorem. In
this small note, I have reconstructed and reformulated the original
argument, thereby resolving the gap, too. As noted by Ghorpade and
Ludhani - what is written literally does not prove, what it claims
to prove. However, the idea is clear, and in this note I will explain
what the original idea in the proof was, and show that it still holds.
This does not change the fact, that the correction given by S. Ghorpade
and R. Ludhani is the straightforward and obvious way to rectify the
gap. 

\section*{Recap of Theorem 1}

Recapitulating Theorem 1 from {[}4{]} using the original notation.\\

Theorem 1: The projective Reed-Muller code $PC_{\nu}(m,q),1\leq\nu\leq m(q-1)$
is an $\left[n,k,d\right]$-code with

\[
n=\frac{q^{m+1}-1}{q-1},\:
\]

\[
k=\mathop{\sum_{\begin{subarray}{c}
t\,=\,\nu\,mod\,q-1\\
0\,<\,t\,\leq\,\nu
\end{subarray}}\left(\sum_{j\,=\,0}^{m\,+\,1}(-1)^{j}\binom{m+1}{j}\binom{t-jq+m}{t-jq}\right),}
\]

\[
d=(q-s)q^{m-r-1},
\]

where $\upsilon-1=r(q-1)+s,\;0\leq s<q-1.$\\

Proof: Use induction on $m.$ For the case $m=1$ see {[}4{]}. 

In the induction step we consider to cases 1) $\upsilon-1=(m-1)(q-1)+s,\;0\leq s<q-1$
and 2) $\upsilon\leq(m-1)(q-1)$. 

For case 1) the argument goes like this:

If $\upsilon-1=(m-1)(q-1)+s,\;0\leq s<q-1$ we would like to prove
\begin{equation}
\left|Z(F)_{\mathbb{F}_{q}}\right|=\frac{q^{m+1}-1}{q-1}\qquad or\qquad\left|Z(F)_{\mathbb{F}_{q}}\right|\leq\frac{q^{m+1}-1}{q-1}-(q-s).
\end{equation}

Assume that (1) is false, i.e. that $0<\left|\mathbb{P}^{m}(\mathbb{F}_{q})\setminus X\right|=t<q-s$,
and let $\mathbb{P}^{m}(\mathbb{F}_{q})=\left\{ P_{1},P_{2},...,P_{t}\right\} $.
Let $G_{i}(X),i=1,...,t-1$ be linear polynomials defining $t-1$
hyperplanes such that $G_{i}(P_{j})=\delta_{ij},i=1,...,t-1,j=1,...,t$.
Then the polynomial $H(X)=F(X)\prod_{i\,=\,1}^{t\,-\,1}G_{i}(X)\;$has
degree $(m-1)(q-1)+s+t\leq m(q-1)$ and $Z(H)_{\mathbb{F}_{q}}=$
$\mathbb{P}^{m}(\mathbb{F}_{q})\setminus\left\{ P_{t}\right\} .$
This contradicts Lemma 4 in {[}3{]}.

See {[}3{]} for case 2).

In the above argument the sentence ``Let $G_{i}(X),i=1,...,t-1$
be linear polynomials defining $t-1$ hyperplanes such that $G_{i}(P_{j})=\delta_{ij},i=1,...,t-1,j=1,...,t$.''
does not imply that $H(X)=F(X)\prod_{i\,=\,1}^{t\,-\,1}G_{i}(X)\;$is
zero in $\left\{ P_{1},P_{2},...,P_{t-1}\right\} .$ This is the gap
found by S.Ghorpade and R.Ludhani. As Ghorpade and Radhani explain
in {[}1, Remark 3.4{]}, it is enough to find hyperplanes $H_{i}$
such that $H_{i}$ contains $P_{i}$, but not $P_{t}$, for $i=1,...,t-1.$
Let $G_{i}(X)$ be the defining polynomial for each $H_{i}$, so $G_{i}(P_{i})=0$
and $G_{i}(P_{t})\neq0$. Now $H(X)=F(X)\prod_{i\,=\,1}^{t\,-\,1}G_{i}(X)\;$actually
have the property $H(P_{i})=0,i=1,...,t-1$ and $H(P_{t})\neq0$.
And it is indeed possible to find these hyperplanes $H_{i}$.\linebreak{}

\section*{Closing the Gap.}

The original idea with the linear polynomials $G_{i}(X),i=1,...,t-1$
was the following: Let $G_{i}(X),i=1,...,t-1$ be linear polynomilas,
that defines $t-1$ hyperplanes $H_{i}$ such that $H_{i}$ contains
$P_{i}$, but not any $P_{j}$, for $i=1,...,t-1,j=1,...,t.$ In other
words $P_{i}\in H_{i}$ and $P_{i}\notin H_{j}\:$ for $i\neq j,i=1,...,t-1\;j=1,...,t$.
According to Lemma 1 below, we can always find linear polynomials
that meet this condition. As before the polynomial $H(X)$ has the
property $H(P)=0\;$for all $P\in\mathbb{P}^{m}(\mathbb{F}_{q})\setminus\left\{ P_{t}\right\} ,$
and $H(P_{t})\neq0.\;$

To put it in another way, we ask for linear polynomials $G_{i}(X),i=1,...,t-1$
such that
\[
G_{i}(P_{j})=\begin{cases}
0 & \textrm{for }i=j\\
\lambda\neq0 & \textrm{for }i\neq j
\end{cases}
\]
 where $i=1,...,t-1,j=1,...,t$. The essence of S.Ghorpade and R.Ludhanis
fix is only to require $G_{i}(P_{i})$=$0$ and $G_{i}(P_{t})$=$\lambda\neq0,i=1,...,t-1$.
\\

Lemma 1: Given $t$ arbitrary points, say $\left\{ P_{1},P_{2},...,P_{t}\right\} ,t<q$
in $\mathbb{P}^{m}(\mathbb{F}_{q})$ then it is always possible to
find hyperplanes $H_{i}$ such that $H_{i}$ contains $P_{i}$, but
not any $P_{j}$ i.e.$P_{i}\in H_{i}$ and $P_{i}\notin H_{j}\:for\;i\neq j,i,j=1,...,t$.\\

Proof: Without loss of generality, it is enough to show, that we always
can find a hyperplane $H_{1}$such that $P_{1}\in H_{i}$and $P_{i}\notin H_{1},j=2,...,t$.

If $m=2$ this is almost trivial, since all $\mathbb{P}^{2}(F_{q})$
is covered by $q+1$ lines through $P_{1}.$ Now the $t-1$ points
$\left\{ P_{2},...,P_{t}\right\} $can only be on maximum $t-1$ different
lines, so choose one of the remaining minimum $q-t+2$ lines ($q-t+2>0$).
That line will define $H_{1}$.

Let $m>2$. The construction of the hyperplane $H_{1}$follow $m-1$
steps.

Step 1: Let $\mathbb{P}^{m}(F_{q})$ be covered by $\frac{q^{m}-1}{q-1}$
lines through $P_{1}.$(this is a well-known fact - see remark 2).
Choose a line $\textrm{\ensuremath{\ell}}$ from the covering such
that$P_{1}\in\ell$ and $P_{i}\notin\ell\:for\;j=2,...,t$. Since
$t-1<q-1<\frac{q^{m}-1}{q-1}$, there is plenty af lines to choose
from.

Step 2: Let $\mathbb{P}^{m}(F_{q})$ be covered by $\frac{q^{m-1}-1}{q-1}$
planes through $\textrm{\ensuremath{\ell}}$ from step 1. Choose a
plane $\wp$ from the covering such that$P_{1}\in\wp$ and $P_{j}\notin\wp\:for\;j=2,...,t$.
Since $t-1<q-1<\frac{q^{m-1}-1}{q-1}$, there is planes to choose
from.

Step $k$: Let $\mathbb{P}^{m}(F_{q})$ be covered by $\frac{q^{m-k+1}-1}{q-1}$
flats of dimension $k$ through the $k-1$-dimensional flat $\mathscr{V}_{k-1}$
from step $k-1$. Choose a flat $\mathscr{V}_{k}$ from the covering
such that that$P_{1}\in\mathscr{V}_{k}$ and $P_{j}\notin\mathscr{V}_{k}\:for\;j=2,...,t$.
Since $t-1<q-1<\frac{q^{m-k+1}-1}{q-1}$, there dimension $k$ flats
to choose from.

(Last) Step $m-1$: Let $\mathbb{P}^{m}(\mathbb{F}_{q})$ be covered
by $\frac{q^{m-(m-1)+1}-1}{q-1}=\frac{q^{2}-1}{q-1}=q+1$ hyperplanes
of dimension $(m-1)$ through the $(m-2)$-dimensional flat $\mathscr{V}_{m-2}$
from step $m-2$. Choose a $(m-1)$-dimensional hyperplane $\mathscr{V}_{m-1}$
from the covering such that that$P_{1}\in\mathscr{V}_{m-1}$ and $P_{j}\notin\mathscr{V}_{m-1}\:for\;j=2,...,t$.
Since $t-1<q-1<q+1$, there is a hyperplane to choose from. The chosen
hyperplane $\mathscr{V}_{m-1}$will do as $H_{1}$, and we are done.\\

Remark 1. To prove Lemma 1, it is not enough to consider an arbitrary
covering of $\mathbb{P}^{m}(\mathbb{F}_{q})$ with $q+1$ hyperplanes
through a common $(m-2)$-dimensional space containing $P_{1}$, and
then use the counting argument to find the hyperplane containing $P_{1}$,
and not $\left\{ P_{2},...,P_{t}\right\} $. One must exclude the
case where some of $\left\{ P_{2},...,P_{t}\right\} $are in the common
$(m-2)$-dimensional space in the covering. The construction used
in the proof of lemma 1 delivers such a covering.\\

Remark 2. It is known that through a given point in $\mathbb{P}^{m}(\mathbb{F}_{q})$
there is $\frac{q^{m}-1}{q-1}$ lines through that point and they
cover all points of $\mathbb{P}^{m}(\mathbb{F}_{q})$. More generally
given a $k$-dimensional flat in $\mathbb{P}^{m}(\mathbb{F}_{q})$
there is covering of $(k+1)$-dimensional flats with the given $k$-dimensional
flat in common and the covering consist of $\frac{q^{m-k+1}-1}{q-1}$
flats of dimension $k$. This can be seen in my Ph.D.Thesis in the
article \textquotedblleft On the number of rational points on codimension-1
algebraic sets in $\mathbb{P}^{n}(\mathbb{F}_{q})$'' ({[}3, p. 24{]}),
where it is stated even in more generality as the number of $r$-dimensional
flats through a fixed $s$-dimensional flat in $\mathbb{P}^{n}(\mathbb{F}_{q}),$
or in {[}2{]}. \\

Remark 3. From the construction in the proof it is clear that Lemma
1 is true even for $t<q+2$.

\end{document}